# Influence of Gold-Selenium Precursor Ratio on Synthesis and Structural Stability of α- and β- AuSe

Aditya Kumar Sahu[1,*] and Satyabrata Raj[1,2]

[1]Department of Physical Sciences, Indian Institute of Science Education and Research Kolkata,
Mohanpur, Nadia, 741246, India

[2]National Centre for High-Pressure Studies, Indian Institute of Science Education and Research Kolkata, Mohanpur, Nadia, 741246, India

[*]Email: aks16rs023@iiserkol.ac.in

## Abstract

Gold selenide (AuSe) is a multilayer compound yet to be thoroughly studied. The colloidal synthesis and characterization of gold selenide nanoparticles are described, emphasizing the effect of different gold-to-selenium precursor ratios and temperatures on the crystal structure and form. The structural characterization is done using an X-ray diffraction pattern. The coexistence of the α- and β-AuSe phases is observed in all synthesized samples. The morphologies of the mainly α-AuSe sample are nanobelts, whereas the primarily β-AuSe phase sample has a nanoplate-like structure, according to the TEM and SEM data. All of the samples had Raman vibrational modes with mixed phases. The effect of high pressure on as-prepared AuSe samples has been studied in this work. The introduction of external pressure and temperature allows both phases to transition. Pressure lowers the existence of other phase modes, and the corresponding dominating sample modes are entirely significant in our sample. The phase transition pressure was observed using Raman scattering. Our findings show that 2D AuSe has a lot of promise for multifunctional applications, encouraging more research on these systems.

# 1. Introduction

Characteristic minerals of noble metals include gold sulfide, selenide, and telluride. [1] The phase relationship of these compounds makes them desirable for a wide range of possible applications, including sensing devices, catalyst supports, energy conversion and storage systems, optoelectronic devices, and solar cells. [2-9] While phase characteristics in Au-Te are well understood, the Au-Se system has received less research. AuSe has been reported to occur in two crystalline phases in the Au-Se system: α-AuSe and β-AuSe. Rabenau et al. [10] reported the crystal structures of α and β-AuSe, while Cretier and Wiegers [11] improved the data, suggesting the system as mixed-valence $(Au^{3+}, Au^{+})Se^{2-}$. Based on thermodynamic research, the current study provides a more thorough analysis of the characteristics of AuSe variations.

The Au/Se mole ratios, time, and temperature are applied to impact the synthesis of α and β phases of AuSe. Due to the layered structure of AuSe, it is possible to construct 2D nanostructures with specific thicknesses and characteristics. [12] Exfoliation is a popular top-down synthetic process for creating 2D nanomaterials from 3D layered structures.[13] Exfoliation produces nanosheets with thicknesses as thin as one atom, but these nanosheets are frequently broken, making production scaling challenges.[13,14] Others have reported on deposition techniques to develop 2D nanomaterials. [15-17] Deposition processes often necessitate costly equipment and operate under high temperatures. [18] Colloidal synthesis is a bottom-up method for developing 2D nanomaterials from chemical precursors. [19] Colloidal synthesis requires relatively mild conditions, low-cost equipment, and feasible scaling-up. [20] The reaction conditions, including time, temperature, concentration, precursor type, and surfactant type, may all be consistently modified, controlling the structure and morphology of the produced nanomaterials, resulting in tunable properties. [21-22]

Raman scattering is one of the most used methods for obtaining information on the lattice vibration of a crystal. [23-25] Raman spectroscopy can even be employed under high pressure to understand the properties of the system. As a result, they have become a popular technological method in material science. Hydrostatic pressure can affect the structural and electronic properties of nanomaterials through volume compression and lattice distortion. A significant strain exists on the surface of the nanocrystalline material due to the enormous surface-to-volume ratio. As a result, high-pressure research is necessary to comprehend its behavior under high-stress situations. The article described the colloidal synthesis and characterization of gold selenide nanomaterial, focusing on the influence of various gold-to-selenium precursor ratios and temperatures on the crystal structure and its shape. Raman spectroscopy is also used to examine the anisotropic vibrational behavior of AuSe. Our research aims to investigate the fundamental characteristics of AuSe to get a complete knowledge of previously undiscovered aspects, which will serve as a foundation for future experimental and theoretical research that will improve future applications. These findings indicate AuSe as a new family of 2D materials and a potential alternative for multifunctional optoelectronics.

## 2. Experimental

### 2.1. Materials synthesis

Gold chloride (99%), elemental selenium powder (99%), oleylamine (OLA 70%), toluene, and absolute ethanol (99.9%) were obtained from Sigma Aldrich and used without further processing.

In a three-neck round bottom flask, argon gas was passed for approximately 30 minutes to achieve the inert conditions before adding the precursors and continuing until the synthesis reaction process was completed. The following synthesis method is adopted in one sample (named A). Initially, 6 ml OLA was heated to 100°C under high magnetic stirring conditions and reflux, then the solution of selenium dissolved in 2 ml OLA was added drop-wise, which turned the solution pale grey. The temperature was then raised to 225°C; at that time, the solution of gold dissolved in 2 ml OLA was added, making the reaction mixture black. At 225°C, the reaction was performed for 4 hours. In another sample (named B) synthesis, in the same inert condition, 6 ml OLA was heated to 100°C under high magnetic stirring conditions and reflux. Then, a solution of gold was dissolved in 2 ml OLA, which turned the solution orange. Then the temperature was raised to 225°C, and the selenium solution was dissolved in 2 ml; OLA was added to it, which made the reaction mixture black. At 225°C, the reaction was performed for 4 hours. Following that, sample synthesis processes were completed, collected, and left to cool naturally. After that, ethanol was added to flocculate the particles and wash away any impurities. After that, the materials were collected using centrifugation at 5500 rpm, and the outputs were washed multiple times before being dried at room temperature. Besides α- and β- forms, major elemental gold and selenium were impurities in our samples.

We have carried out synthesis in various temperatures starting from ambient to 340°C and analyzed the final AuSe product to understand the content of Au and Se in the sample. Later, the synthesized samples were reacted in sealed quartz tubes to create the samples. After the initial reaction, samples were annealed at 250-350°C various times (24 hours to 48 hours). The samples are extracted from the tube and grinded with a mortar pestle in the air medium. Manually controlled heating rates of 3°C/min and 5°C/min cooling rates have been fixed. Temperatures at the melting points of high-purity elements were used to calibrate the reaction temperature. Samples A and B were heated to 280˚C and 340˚C, respectively, in our synthesis procedure.

## 2.2. Sample Characterization:

The powder XRD of the AuSe material has been measured with an x-ray diffractometer having monochromatic Cu Kα radiation at 40 kV and 15 mA. Measurements were taken for 2θ values over 10-90˚ in steps of 0.02˚ and at room temperature of 294 K. Raman spectroscopy is carried out by a spectrometer of HORIBA. JEM-2100P microscope was used for transmission electron microscopy (TEM). The samples were first suspended in toluene, then a drop of the suspended nanomaterial was dropped upon a copper grid. Before analysis, the grid with the sample was allowed to dry at room temperature. The field emission scanning electron microscope (FE-SEM) image of AuSe samples was obtained with Carl Zeiss Field Emission Scanning Electron Microscope. The powder samples were suspended in ethanol for imaging and deposited onto ultrasonically cleaned Si (100) substrates using a spin coater. The high-pressure Raman measurements were carried out at room temperature under hydrostatic pressure in a gasket diamond anvil cell (DAC). A DAC is used to produce high pressure on the sample. The sample chamber was created by drilling a hole in a stainless steel gasket with a diameter of 100 μm. A little powder sample and ruby crystals were inserted into the hole of the stainless steel gasket. We use a 4:1 methanol-ethanol combination as a pressure medium. The high pressure is calculated from the frequency shift of the ruby crystals. The Raman spectra were acquired using a 532 nm laser and a 50× objective lens on an Olympus microscope.

## 3. Results and Discussion

The morphology, phase structure, and crystal size of the samples were examined and quantified using X-ray diffraction (XRD), transmission electron microscope (TEM), field emission scanning electron microscope (FESEM), Raman, energy dispersive X-ray spectroscopy (EDS), and chemical element mapping. Colloidal synthesis is advantageous because it allows for the fabrication of nanocrystals with controlled size and shape. Different factors, including time, temperature, solvent, and starting materials, can be changed to produce desired nanoparticles. In the reaction, selenium nuclei are initially developed, and α-AuSe is formed when the gold precursor is added above 200 °C. Following that, the concentration of selenium ions reduces, favoring the synthesis of β-AuSe due to an overabundance of gold ions. The excess gold nuclei subsequently form gold nanoparticles resulting in the high-intensity gold peaks as exhibited. On the other hand, adding the gold precursor at low temperatures produces a gold-rich environment that favors the synthesis of β-AuSe, as Rabenau et al. previously reported. [26] However, there are still minor residues of gold in the reaction; Machogo et al. have previously demonstrated that gold persists independent of the Au:Se ratio. [27] In addition, the annealing results significantly impact the formation of the compounds.

The crystal phase and purity of the synthesized materials were determined by x-ray diffraction. The room temperature diffraction data of powered samples revealed the presence of two intermediate phases. The relative proportions of the two phases were affected by the composition and the temperature condition of the synthesis process. Temperature studies can be used to influence and change crystal phases. Temperature variation might also have an impact on the purity of the samples. We compared our resulting data with the available diffraction data of α-AuSe (JCPDS: PDF#20-0457) and β-AuSe (JCPDS: PDF#20-0458). We found that when Au/Se compositions were put in a furnace at ambient temperature and steadily heated to 280° C, the sample outcomes as a mixture of both phases and the α-phase was the dominant phase. When the mixture was heated to 340° C, the β-phase was formed, having small α-phase proportions, as shown in *Figure 1*. Although the percentages of the two phases varied depending on the approach and reaction temperature, the initial outcome of the synthesis was unpredictable. According to our study, the β-phase appeared mostly in gold-rich synthesis, while the α-phase appears to be in the selenium-rich synthesis. The presence of α and β- AuSe phases indicates that the reaction medium is favorable for forming both the crystal growing processes. This suggests that both phases have very close formation energy.

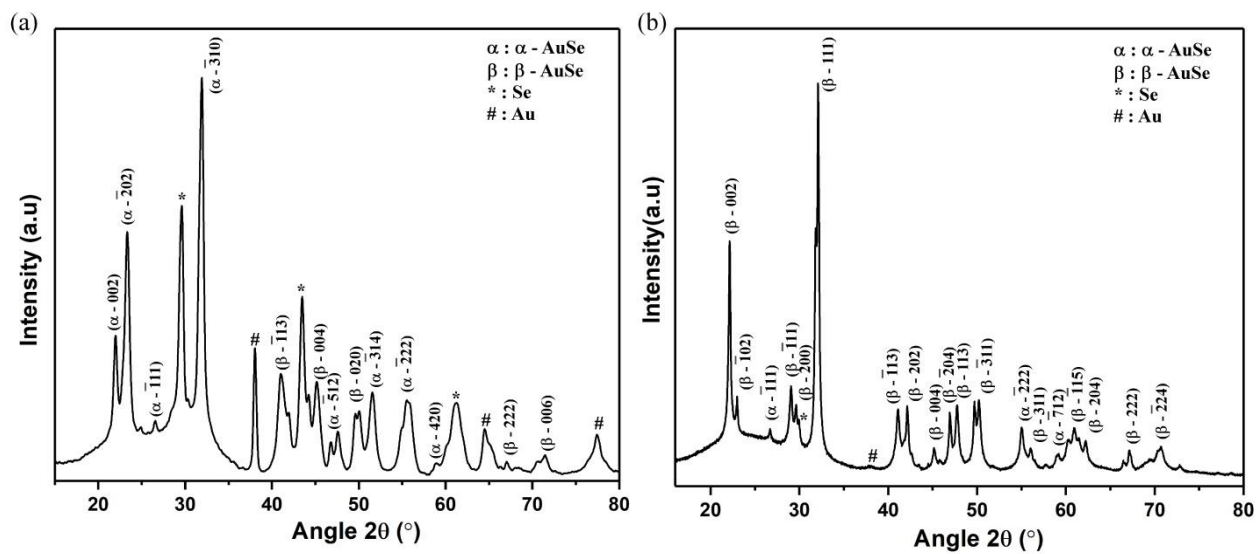

*Figure 1.* XRD patterns of AuSe synthesized in different methods (a) Sample A, (b) Sample B.

The difficulty in attaining thermal equilibrium between the two gold selenide phases at the temperature involved makes studying their phase relationships difficult. However, the stability of the phase described by studies has not been established within the experimental limits of analysis. The formation temperature of both phases is small, and as a result, the driving force for the transformation from each other is also small. Experiments on colloidal synthesis provide unclear findings as well. α and β phase is only observed in crystals under a high excess of selenium and gold impurities. Although there is a chance that the metastable phase would arise, there is no direct evidence, and more investigation into this problem is still an open to this subject.

Electron microscopy was used to investigate the morphological properties of the nanomaterials. Transmission electron microscope (TEM) images of both phases of AuSe are shown in *Figure 2*. Lightly contrasting AuSe nanobelts of varying lengths and widths, as shown in *Figure 2(a),* are related to α-AuSe. *Figure 2 (b)* indicates significantly contrasted larger AuSe nanoplates of varying sizes: β-AuSe. Some nanoplates had more cylindrical edges, while others were hexagonal with distinct facets. The lattice fringes on these gold selenide nanobelts are exceedingly clear, indicating their crystalline structure. Thus temperature experiments can be used to influence and change crystal phases. Furthermore, temperature changes may also impact the purity of the products. Hence we may be sure that the sample prepared at 280 ˚C is dominated by α-phase whereas the sample prepared at 340˚C is dominated by β-phase.

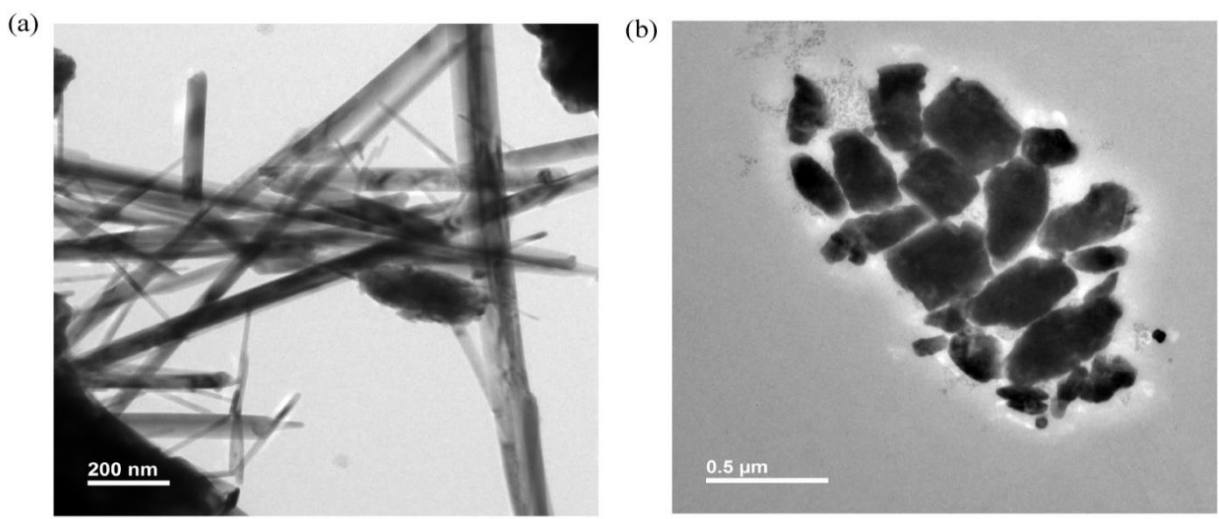

*Figure 2.* TEM images of the AuSe nanocrystals using different reaction methods (a) sample A dominated by α- phase and (b) sample B dominated by β-phase.

Energy Dispersive X-Ray Analysis (EDX) systems are add-ons to Transmission Electron Microscopy (TEM) that use the imaging potential of the microscope to determine the presence of elements in the sample of interest. TEM EDX mapping was used to determine the extent of homogeneity in the samples, and the findings are displayed in *Figure 3*. Column (i) indicates HAADF STEM pictures of the samples employed as focus zones for each sample. Elemental mapping analysis shows the distribution of Au and Se on the nanocrystals in columns (ii) and (iii), respectively, for both samples. The concentration of selenium ions reduces in the β-AuSe dominance sample. As discussed, the β-phase appeared mostly in gold-rich mixtures, while the α-phase appeared in selenium-rich combinations.

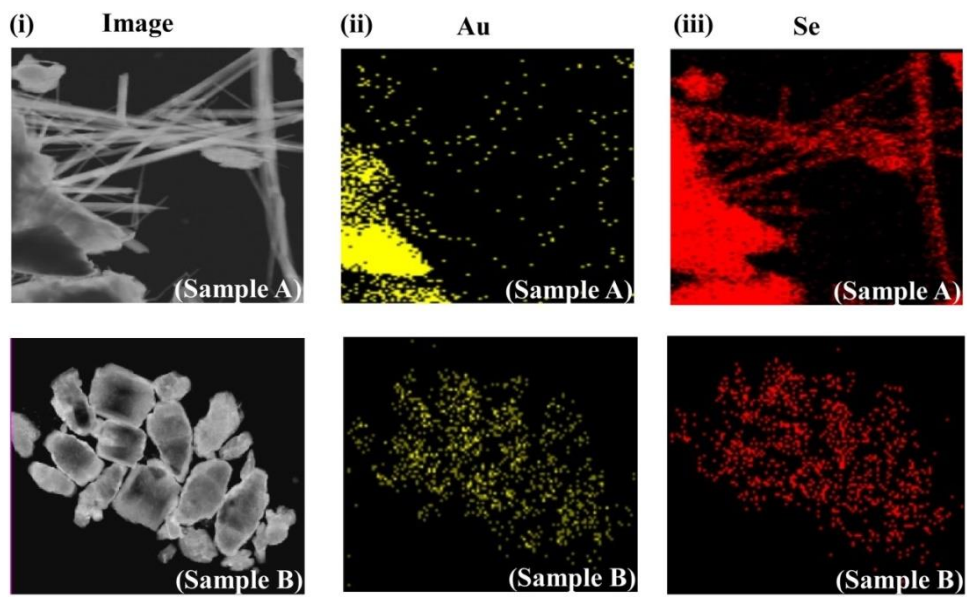

***Figure 3.*** *(i) TEM mapping images of the samples (A) α-AuSe and (B) β-AuSe. Elemental mapping of (ii) Au and (iii) Se in samples A and B, respectively.*

The SEM pictures of the synthesized materials are shown in *Figure 4*. Nanobelts structures were formed in the α-AuSe samples, whereas nanoplates were formed for β-AuSe, as previously established from TEM images.

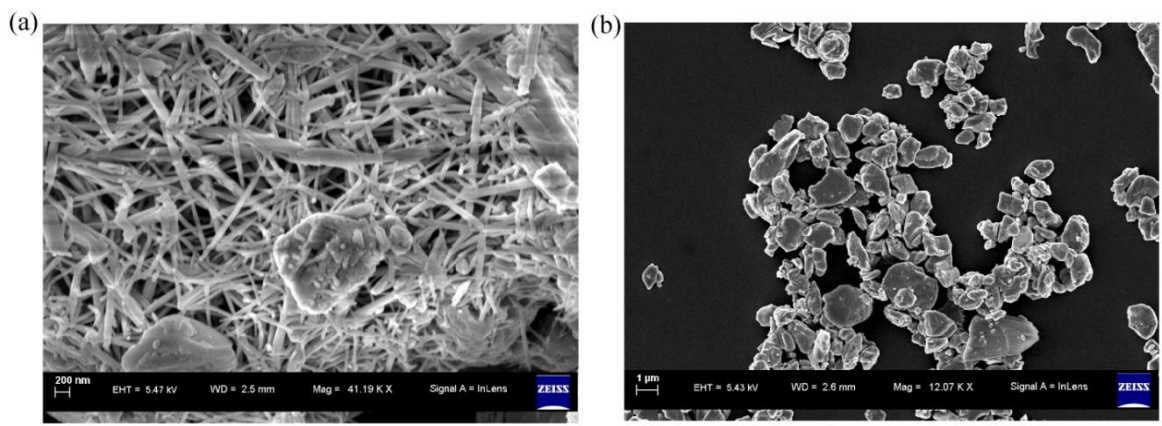

***Figure 4.*** *SEM images of AuSe samples for different phases (a) sample A with dominated α-phase and (b) sample B with dominated β-phase.*

The structure of gold selenide was studied further using a Raman spectrometer. As shown in *Figure 5*, the experimental Raman spectra of sample A (heated to 280°C, i.e., α-AuSe rich) show four main peaks. In contrast, the experimental Raman spectra of sample B (heated to 340°C, i.e., β-AuSe rich) show three major peaks, which agree with prior theoretical and experimental results. [28,29] The Raman peaks were detected at 143.7 cm$^{-1}$, 171.7 cm$^{-1}$, 205 cm$^{-1}$, and 237.4 cm$^{-1}$ in sample A. The strong shift at 237.4 cm$^{-1}$ corresponds to the α-AuSe. The shift at 171.7 cm$^{-1}$ is related to B$_g$ normal mode planar bending, whereas the shift at 237.4 cm$^{-1}$ is A$_g$ normal mode due to the symmetric stretching of the Au-Se bond. The peak at 205 cm$^{-1}$ corresponds to β -AuSe. The Raman spectra of the sample heated to 340° C exhibited vibrational peaks at 172 cm$^{-1}$, 206.2 cm$^{-1}$, and 233 cm$^{-1}$. The peak at 172 cm$^{-1}$ corresponds to planar bending, and at 206.2 cm$^{-1}$ corresponds to the symmetric stretching vibration (A$_g$) of the β-AuSe phase. The intensity of the B$_g$ peak is significantly higher than that of the A$_g$ peak, suggesting that the B$_g$ mode is dominant. Both modes vibrate inside the layered plane, and no vibration mode is observed perpendicular to the plane. The results demonstrate that the proposed samples comprise the α- and β-AuSe phases. Though sample A contains both the α-AuSe and β-AuSe peaks at ambient conditions, the α-AuSe phase is the dominant phase, whereas sample B contains predominant β-AuSe peaks at ambient conditions.

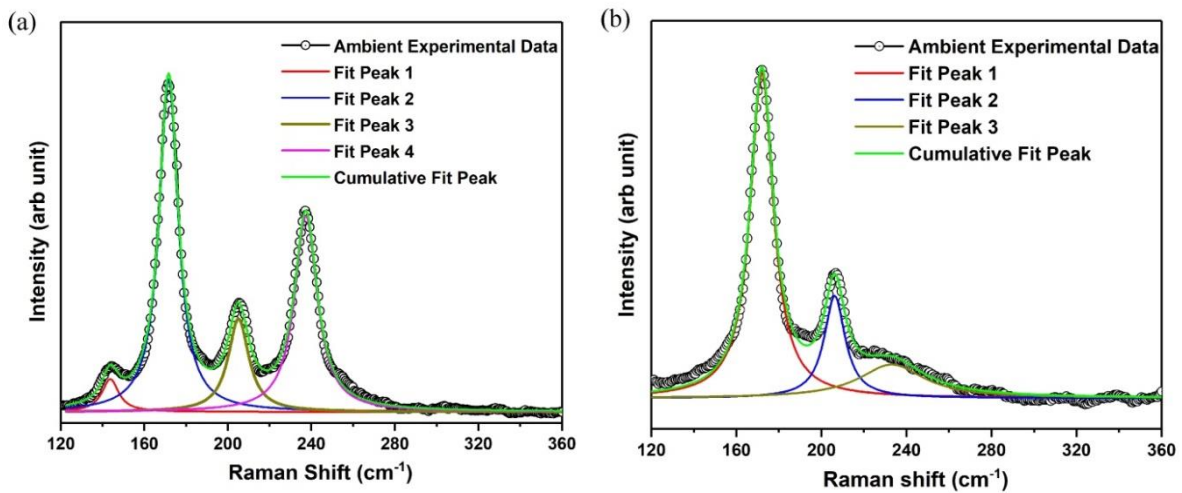

*Figure 5.* *Raman modes of (a) sample A and (b) sample B in the ambient environment*

Raman spectroscopy is an effective technique for identifying pressure-induced structural phase transitions in solids, which generally appear as abrupt changes in vibration mode behavior. For example, symmetry lowering caused by phase transitions usually leads to Raman peak splitting. The initial symmetry of the crystal is generally lower, and there are substantial Raman modes. At the same time, more forms of competing for interatomic and intermolecular interactions (such as hydrogen bonds, van der Waals forces, electrostatic interactions, and charge transfer) influence the vibration force constants. The equilibrium between these different interactions shifts at high pressure, potentially causing Raman mode behavior discontinuity.

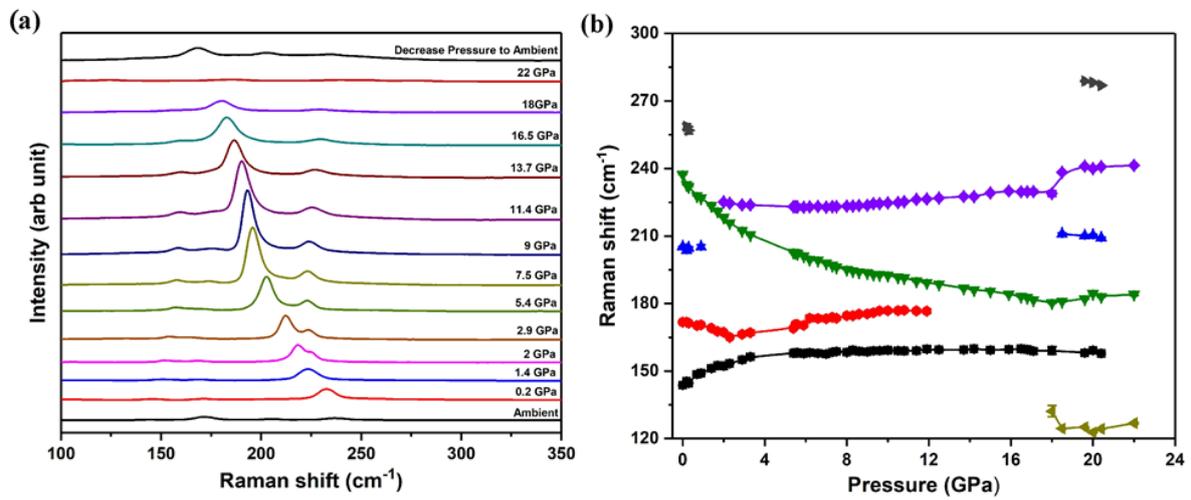

*Figure 6. (a) High-pressure Raman spectra of AuSe sample A at various pressure and (b) Raman shifts of the significant Raman modes with increasing pressure. The connecting lines are a reference for detecting changes in the Raman shift.*

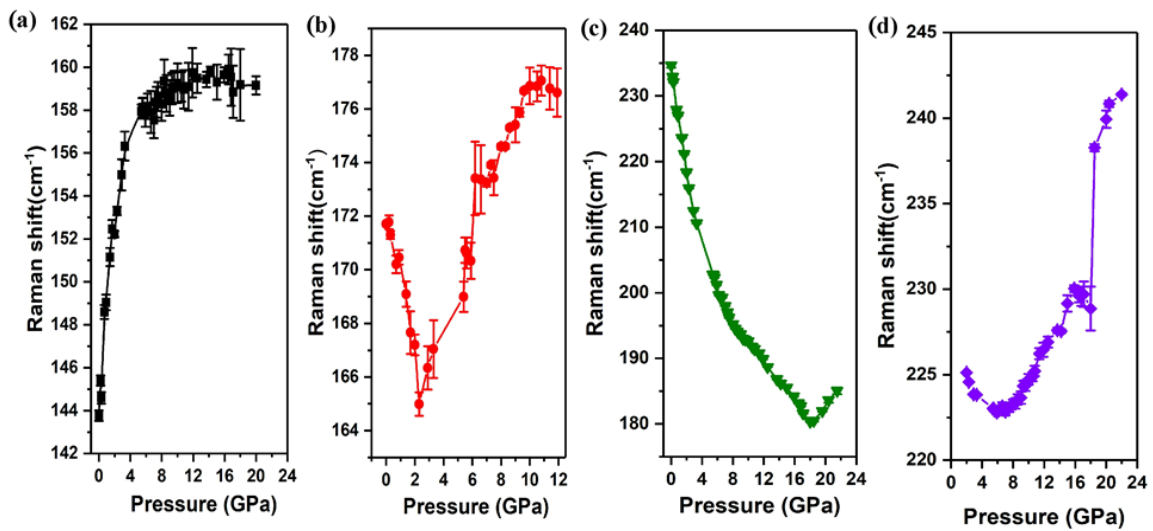

*Figure 7. The pressure behavior of various Raman modes of sample-A (a)-(d). The connecting lines are a reference for detecting changes in the Raman shift.*

High-pressure Raman spectroscopy has been carried out on these samples. When the pressure of 0.2 GPa pressure is applied to sample A, the ambient peak intensity around

171.7 cm$^{-1}$ decreases, but the peak intensity at 237.4 cm$^{-1}$ increases and becomes the most intense with pressure, as shown in *Figure 6(a)*. This indicates that the presence of β modes reduces on applying pressure, and α modes are significant due to the low ground state energy difference between these two phases. Both peaks around 171.7 cm$^{-1}$ and 237.4 cm$^{-1}$ are redshifted when the pressure is applied initially, as shown in *Figure 6 (b)*. The peaks at lower wavenumbers are broad and asymmetric, but those at higher wavenumbers are noticeably sharp and symmetric, as seen in *Figure 6*. A moderate intensity peak was noticed at increased pressures around 205 cm$^{-1}$, which diminished when the pressure reached 1 GPa. These peaks broaden and diminish with increased pressure and decreasing intermolecular distances. The peak of about 265 cm$^{-1}$ is due to the antisymmetric stretching mode that further dissipates with growing pressure. *Figure 7 (a)-(d)* indicates the relative shifts of the primary modes. When the pressure is increased to 20 GPa, the peak around 144 cm$^{-1}$ blueshifts to 159 cm$^{-1}$. Peak splitting of the intense peak occurs at a pressure of 2 GPa. This reflects a shift in the inter-layer molecular interaction. At that same 2GPa, the peak around 170 cm$^{-1}$ redshifted initially became blueshifted as pressure increased, and the peak disappeared when the pressure reached 12 GPa. At 2 GPa, the intense peak split into two peaks with shifts of 218.3 cm$^{-1}$ and 225.1 cm$^{-1}$. With increasing pressure, the peak of 218.3 cm$^{-1}$ redshifted to 180.33 cm$^{-1}$ up to 18 GPa and blueshifted to 184.1 cm-1 at 22 GPa. The Raman intensity decreases as the molecule becomes dense over 10 GPa, indicating electronic/structural changes. Up to 7 GPa, the second splitting peak of 225 cm$^{-1}$ redshifted to 223 cm$^{-1}$ and then blueshifted when pressure increased to 22 GPa. Some anomaly is observed in the sample at 18 GPa. Peak intensity has decreased as pressure increased from 18 GPa to 22 GPa, with the appearance of some more low-intensity peaks around 209-210 cm$^{-1}$ and 273-278 cm$^{-1}$. The formation of new peaks at 124 cm$^{-1}$ - 130 cm$^{-1}$ confirms a shift in the electronic state, resulting in a change in symmetry. At 22 GPa, the peak intensity progressively decreases and vanishes as the pressure rises. We detected three Raman peaks at 168.2 cm$^{-1}$, 202 cm$^{-1}$, and 236 cm$^{-1}$, releasing pressure to ambient conditions. When pressure is released to the ambient, the intense peak is 168.2 cm$^{-1}$.

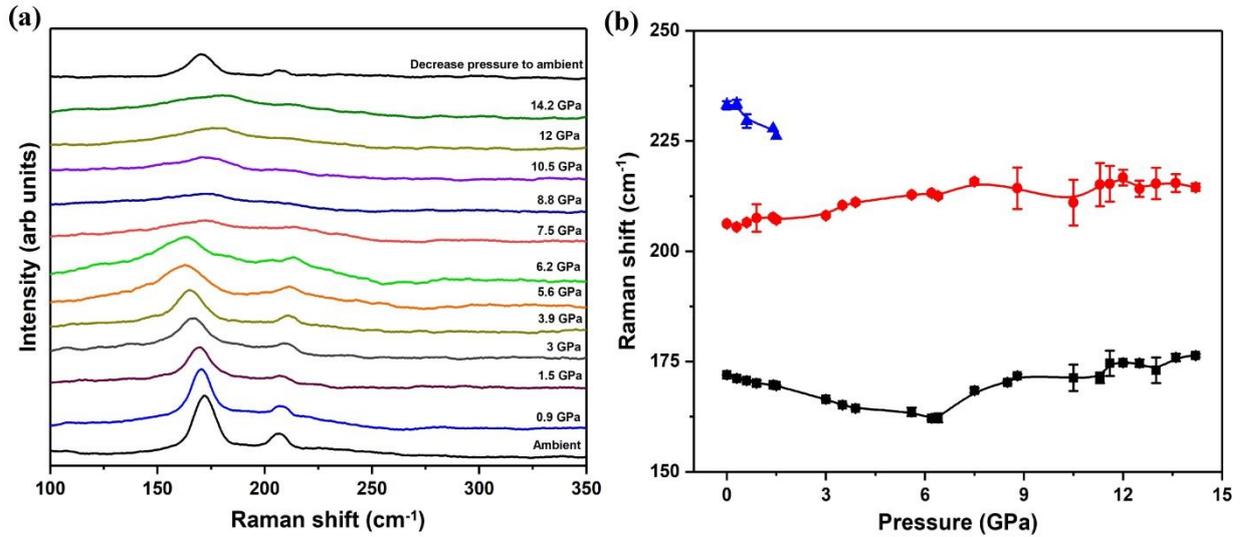

*Figure 8.* *(a) High-pressure Raman spectra of AuSe sample B at various pressure and (b) Raman shifts of the significant Raman modes with increasing pressure. The connecting lines are a reference for detecting changes in the Raman shift.*

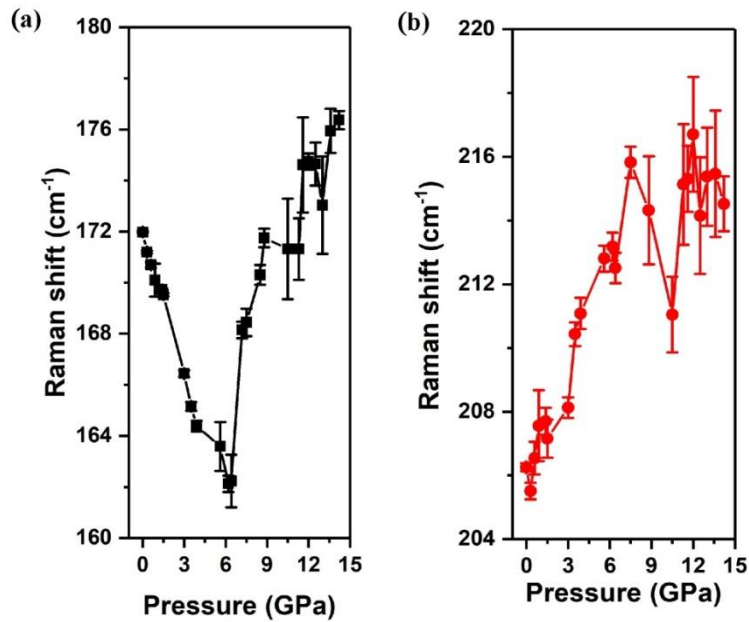

*Figure 9.* *The pressure behavior of various Raman modes of sample-B (a)-(b). The connecting lines are a reference for detecting changes in the Raman shift.*

When pressure is applied to sample B, the peak around 172 cm$^{-1}$ is redshifted to 162.5 cm$^{-1}$ up to 6.5 GPa (Figure 8) and blue-shifted to 176.3 cm$^{-1}$ when pressure increased to 14.2 GPa. Up to 1.5 GPa, some low-intensity peaks develop about 233 to 226 cm$^{-1,}$ but the later peak vanishes. The peak at 206 cm$^{-1}$ blue-shifted to 216 cm$^{-1}$ up to 8.5 GPa and then stayed in the range of experimental error. With rising pressure, the peak becomes broader and flattens at 15 GPa. *Figure 9 (a)-(b)* indicates the relative shifts of the major modes. When we released the pressure to the ambient, we found two significant peaks at 170.2 cm$^{-1}$ and 206.2 cm$^{-1}$. With increased pressure, the maximum frequency reaches 216 cm$^{-1}$. The maximum frequency of the computed phonon spectrum is around 240 cm$^{-1}$ (smaller than 450 cm$^{-1}$ for covalent P-P bonding in black phosphorene and 470 cm$^{-1}$ for ionic Mo-S bonding in MoS$_2$), implying a new form of chemical interaction between Au and Se atoms. 2D β-AuSe exhibits the same in-plane configuration as the bulk, and the number of layers may be increased, resulting in a significant reduction in the thickness of each layer, accompanied by a minor rise in the in-plane lattice constants.

The shifts in the Raman peaks can be caused by a variety of factors, including (i) Peak shifts are seen when layered materials evolve from bulk to 2D monolayers [30], (ii) in the presence of defects such as extrinsic impurities [31], and (iii) lattice strain produced by defects or morphology [32]. Not detecting bands are likely to be weak or hidden by stronger bands. As pressure increases, the Raman bands shift to higher wavenumbers because interatomic and intermolecular interactions strengthen as they compress the system. The corresponding volume compression mainly influences the pressure-induced wavenumber shifts. A structural change in molecules usually causes spectral shifts. Additional Raman peaks may arise as a result of reducing the symmetry. Reduced height and asymmetry in peaks in crystalline compounds indicate decreased crystallinity and an irregular arrangement of atoms in the lattice. Usually, tensile tensions cause Raman modes to redshift. The inhomogeneous broadening is more pronounced in amorphous materials, where the chemical composition, crystal size, molecular chain length, morphology, and local environment differ significantly. Plasmon confinement is responsible for peak broadening in nanocrystalline materials. The results of the experiments showed that both compounds had an anomaly. These findings indicate further investigation that could explain the modes' unique behavior for both compounds. Pressure changes the relative geometry between nuclei, bends the electron clouds, and, as a result, modulates the restoring forces. As a result, for short and linear Au-Se bonds, the repulsive terms of the interatomic potential inhibit the approximation of the Au and Se atoms due to compression, lowering the angle of the bond. We observed that high pressure reduces the existence of other phase modes, and the corresponding dominating sample modes are entirely significant in our sample. So it is possible that both phases can interchange themselves by applying external pressure and temperature.

## 4. Conclusion

This work projected the thermodynamic stability of various phases of the AuSe system produced across a wide temperature range. The morphology, phase structure, and crystal size of the samples were observed and characterized by x-ray diffraction, transmission electron microscopy, high-resolution scanning electron microscopy, Raman spectroscopy, energy-dispersive x-ray spectroscopy, and chemical element mapping. Several parameters, such as time, temperature, solvent, and starting materials, can be modified to make desired nanoparticles. The typical thermodynamic characteristics of gold selenide are observed based on the data acquired in this investigation. The room temperature diffraction data of powder samples revealed the presence of two intermediate phases, α-AuSe and β-AuSe. Electron microscopy shows that α-AuSe has a nanobelts structure, whereas β-AuSe has a nanoplate shape. The Raman mode shift of the α- and β-AuSe was studied under high pressure. We observed that high pressure reduces the occurrence of other phase modes, and the corresponding dominating sample modes are completely significant in our sample. As a result, both transition phases could be possible by adding external pressure and temperature. Nonetheless, AuSe, a novel system that contributes to the 2D family of materials, is utilized in applications as electrode materials in solar cells, batteries, and supercapacitors, comparable to graphene. These advantageous characteristics position the 2D AuSe as a suitable target for device applications. All of the work covered in our studies would be useful for further AuSe material research and applications.